\documentclass[11pt]{article}
\usepackage{graphicx}
\usepackage{amsmath}
\usepackage{amsfonts}
\usepackage{amssymb}
\usepackage{epsfig}
\usepackage{times}
\usepackage{cite}
\usepackage{calc}
\usepackage{version}
\usepackage[english]{babel}

\newcommand{\nc}{N_{_c}}
\newcommand{\cf}{C_{_F}}
\newcommand{\tr}{T_{_R}}
\newcommand{\nf}{n_{f}}
\newcommand{\ns}{N_{s}}
\newcommand{\calp}{{\cal P}}

\begin{document}

\begin{titlepage}

\null

\vskip 1.5cm

\vskip 1.cm

{\bf\large\baselineskip 20pt
\begin{center}
\begin{Large}
Collimation of energy in medium-modified QCD jets
\end{Large}
\end{center}
}
\vskip 1cm

\begin{center}
Redamy P\'erez-Ramos 
\footnote{e-mail: redamy.perez@uv.es}\\
\smallskip
Department of Physics, P.O. Box 35, FI-40014 University of Jyv\"askyl\"a, Finland and\\
\smallskip
Departament de F\'{\i}sica Te\`orica and IFIC, 
Universitat de Val\`encia - CSIC\\
Dr. Moliner 50, E-46100 Burjassot, Spain\\
\medskip
Vincent Mathieu
\footnote{e-mail: vincent.mathieu@umons.ac.be}\\
\smallskip
ECT*, Villa Tambosi, I-38123 Villazzano (Trento), Italy
\end{center}

\baselineskip=15pt

\vskip 3.5cm

{\bf Abstract}: The collimation of energy inside medium-modified jets is 
investigated in the leading logarithmic approximation of QCD.
The Dokshitzer-Gribov-Lipatov-Altarelli-Parisi (DGLAP) evolution equations are
slightly modified by introducing splitting functions enhanced in the infrared sector.
As compared to elementary collisions in the vacuum, the angular distribution of the jet energy
is found to broaden in QCD media.
\end{titlepage}

\section{Introduction}

The jet quenching has been established as the phenomenon
of strong high-$p_T$ suppression of partons produced 
in heavy ion collisions \cite{Adcox:2001jp,Adler:2003ii,dEnterria:2009am}. 
Experimentally, it has been confirmed that highly virtual partons 
produced in such reactions suffer an energy degradation 
prior to their hadronization in the vacuum.
The high-$p_T$ partons energy is radiated away as gluon 
bremsstrahlung, thus increasing the amount of such gluons after several
interactions of the leading parton with the scattering centers in the 
quark gluon plasma (QGP) occur \cite{Baier:2000mf}. Jet production in 
hadronic collisions is a paradigmatic hard QCD process. An
elastic ($2\to2$) or inelastic ($2\to2+X$) scattering of two partons from each one of
the colliding hadrons (or nuclei) results in the production of two or more partons
in the final-state. At high $p_T$ , the outgoing partons have a large virtuality Q which
they reduce by subsequently radiating gluons and/or splitting into quark-antiquark
pairs. Such a parton branching evolution is governed by the QCD branching probabilities
given by the DGLAP equations \cite{Dokshitzer:1978hw} down to virtualities ${\cal O}(1\,\text{GeV}^2)$.

In this paper we study the energy collimation of jets produced in heavy ions collisions
from a toy QCD-inspired model introduced by Borghini and Wiedemann \cite{Borghini:2005em}. 
In this model, the DGLAP splitting functions are enhanced in the infrared region in order to
mimic medium-induced soft gluon radiation. More precisely, the $1/x$ dependence of
the branching probabilities $q\to qg$, $g\to gg$ is enhanced by introducing the factor $N_s=1+f_{med}$ in 
$P_{ab}(x)=(1+f_{med})/x+{\cal O}(1)$. In order to quantify the collimation of energy, we consider a 
jet of half opening angle $\Theta_0$ initiated by a parton $A$, followed by the production 
of a subjet of half opening angle $\Theta$ (with $\Theta<\Theta_0$) initiated by a parton $B$ with 
large energy fraction $x\sim1$, which contains the bulk of the jet energy registered by the calorimeter.
Experimentally, this corresponds to the calorimetric measurement of the energy flux deposited within
a certain solid angle $\Theta$. The first version of this quantification was given in 
\cite{Dokshitzer:1988bq,Dokshitzer:1991wu} in the vacuum and it was found to scale like,
$$
\frac{\Theta}{\Theta_0}=\left(\frac{E\Theta_0}{\Lambda}\right)^{-\gamma_A(x)},
$$
where $\gamma_A(x)$ ($A=q,\bar q,g$) is a certain function derived from the DGLAP evolution equations at 
large $x$ and $\Lambda\equiv\Lambda_{QCD}$ is the mass scale of QCD. In this context, we will introduce 
the $N_s$ dependence in $\gamma(x)\to\gamma(x,N_s)$ in order to compare our results in the medium 
($\gamma(x,N_s)$) with previous results in the vacuum ($\gamma(x)=\gamma(x,1)$). The same logic was applied 
to the study of the collimation of average multiplicities in \cite{Arleo:2009yu}, where jets were found 
to broaden as compared to those in the vacuum after inserting the $N_s$-dependence in the evolution.

The same model has been largely used, i.e. for the estimation of other observables like medium-modified
inclusive $p_T$-distributions \cite{Borghini:2009eq} and fragmentation functions \cite{Albino:2009hu}, 
in all cases results look promising and appealing for future comparison with experimental data. The 
large suppression of these distributions \cite{Albino:2009hu,Borghini:2009eq} at large 
$x$ (high-$p_T$) and the enhancement at small $x$ (small-$p_T$) are respectively good examples 
of the jet quenching in dense QGP.

\section{Theoretical framework}
Let us start from the production of one gluon or quark ($A=g,q,\bar q$) initiated jet 
with energy $E$, followed by its fragmentation into another parton $B$ at angle
$\Theta$ such that, $Q_0/E\leq\Theta\leq\Theta_0$, where $\Theta_0$ is the half opening angle of
the jet and $Q_0$, the minimal transverse momentum of the emitted partons before the
hadronization into hadrons occurs. The collimation of the energy is characterized by the 
energy fraction $x\sim1$ where the bulk of the jet energy inside the
given cone $\Theta<\Theta_0\ll1$ is deposited. The probability for the energy fraction $x$
to be deposited in a cone of aperture $\Theta$ is related to the inclusive spectrum
of partons through the formula \cite{Dokshitzer:1991wu},
\begin{equation}\label{eq:sumDAB}
D_A(x,E\Theta_0,E\Theta)=\sum_{B=g,q}D_A^B(x,E\Theta_0,E\Theta),
\end{equation}
where the nature of partons $B$ is not identified. In the above relation 
(\ref{eq:sumDAB}) two scales $E\Theta_0$ and $E\Theta$ have been written in the
argument of the inclusive spectrum so as to 
account for the given process in the energy range $E\Theta\leq Q\leq E\Theta_0$;
this notation will be shortened below. In order to quantify the collimation of energy, it 
should be considered that the deposited fraction of the jet
energy is large, that is for $x\to1$. 
The DGLAP evolution equations in the vacuum takes the simple form \cite{Dokshitzer:1978hw},
\begin{equation}\label{eq:dglapx}
\frac{d}{d\ln Q^2}D(x,Q^2)=\frac{\alpha_s(Q^2)}{4\pi}\int_x^1\frac{dz}zP(z)D\left(\frac{x}z,Q^2\right),
\end{equation}
where $P$ is the {\em Hamiltonian} matrix of splitting functions characterizing
the parton splitting probabilities. The splitting functions
are accurately obtained from the FO approach in perturbation theory for small coupling 
constant $\alpha_s\ll1$. The coupling constant is written in the form,
$$
\alpha_s(Q^2)=\frac{\pi}{N_c\beta_0\ln(Q^2/\Lambda^2_{QCD})},
$$ 
where $\beta_0=\frac1{4N_c}(\frac{11}3N_c-\frac43T_R)$
is the first coefficient of the beta function and $\Lambda_{QCD}\approx250$ MeV. 
In Mellin space ${\cal D}(j,Q^2)$ is obtained from the transformation
$$
{\cal D}(j,Q^2)=\int_0^1dx\,x^{j-1}D(x,Q^2),
$$
such that the convolution (\ref{eq:dglapx}) simply becomes
$$
\frac{d}{d\ln Q^2}{\cal D}(j,Q^2)={\cal P}(j){\cal D}(j,Q^2),
$$
or more explicitly rewritten in the matrix form at LO
\begin{equation}\label{eq:dglap}
\frac{d}{d\xi}
\begin{pmatrix}
{\cal D}_{q_{\rm NS}}(j,\xi) \\
{\cal D}_{q_{\rm S}}(j,\xi) \\
{\cal D}_{g}(j,\xi)
\end{pmatrix} 
=\begin{pmatrix}
\calp_{qq}(j)&0&0 \\
0&\calp_{qq}(j)&\calp_{q g}(j)\\
0&\calp_{gq}(j)&\calp_{gg}(j) 
\end{pmatrix}
\begin{pmatrix}
{\cal D}_{q_{\rm NS}}(j,\xi) \\
{\cal D}_{q_{\rm S}}(j,\xi) \\
{\cal D}_{g}(j,\xi)
\end{pmatrix},
\end{equation}
where ${\cal D}_{q_{\rm NS}}$ and ${\cal D}_{q_{\rm S}}$ 
stand respectively for the flavor non-singlet (or valence) 
and flavor-singlet quark distributions, and $\calp_{ik}(j)$ 
is the Mellin transform of the LO splitting functions. 
The variable \cite{Dokshitzer:1978hw}
$$
\xi(Q^2)=\frac1{4N_c\beta_0}\ln\left(\ln\frac{Q^2}{\Lambda^2_{QCD}}\right),\quad
\frac{d}{d\ln Q^2}=\frac{\alpha_s}{4\pi}\frac{d}{d\xi},
$$ 
is introduced for the sake of simplicity. In order to account for the medium-induced gluon 
radiation in heavy-ion collisions, we make use of the QCD-inspired model 
proposed in~\cite{Borghini:2005em}, which allows for 
a simple computation of the equations at large $x$. In this model the infrared 
parts of the splitting functions are enhanced by the factor $\ns=1+f_{med}$,
where $f_{med}>0$ accounts for medium-induced gluon radiation. Realistic values
of $f_{med}$ ($=0.6, 0.8$) are extracted from fits of the nuclear modification factor 
$R_{AA}$ \footnote{$R_{AA}$ corresponds to the ratio of medium-modified and unmodified 
single inclusive hadron spectra.} to the RHIC data \cite{Adler:2003ii}; for the LHC we will
take $f_{med}=1$. The medium-modified splitting functions read,
\begin{subequations}
\begin{equation}\label{eq:splitG}
P_{gg}(z)=4\nc\left[\frac{\ns}{z}+\left[\frac{\ns}{1-z}\right]_+
+z(1-z)-2\right],\quad
P_{gq}(z)=2\ \tr[z^2+(1-z)^2],
\end{equation}
\begin{equation}\label{eq:splitQ}
P_{q g}(z)=2\ \cf\left(\frac{2\ns}{z}+z-2\right),\quad
P_{qq}(z)=2\ \cf\left(\left[\frac{2\ns}{1-z}\right]_+-1
-z\right),
\end{equation}
\end{subequations}
with the $[\dots]_+$ prescription defined as $\int_0^1dx[F(x)]_+g(x)\equiv\int_0^1dxF(x)[g(x)-g(1)]$. 
Performing the Mellin transform of Eq.~(\ref{eq:splitG},\ref{eq:splitQ}) gives~\cite{Albino:2009hu}
\begin{subequations}\label{eq:splitmellin}
\begin{eqnarray}
\calp_{gg}(j)\!&\!=\!&\!-4 \nc\left[\ns\ \psi(j+1)+\ns\gamma_E-\frac{\ns-1}{j}-
\frac{\ns-1}{j-1}\right]\nonumber\\ 
&&+\frac{11 \nc}3-\frac{2\nf}3+\frac{8\nc(j^2+j+1)}{j(j^2-1)(j+2)}\label{eq:nuGbis},\\
\calp_{gq}(j)\!&\!=\!&\!\tr\ \frac{j^2+j+2}{j(j+1)(j+2)},\label{eq:splitjG} \\
\calp_{q g}(j)\!&\!=\!&\!2\ \cf\
\frac{(2 \ns-1)(j^2+j)+2}{j(j^2-1)}\label{eq:splitjQ},\\
\calp_{qq}(j)\!&\!=\!&\!-\cf\left[4\ns\ \psi(j+1)+4\ns\gamma_E-4\frac{\ns-1}j
-3-\frac2{j(j+1)}\right]\label{eq:nuFbis},
\end{eqnarray}
\end{subequations}
where $\psi(j)=\frac{d}{dj}\ln\Gamma(j)$ is the digamma function with $\gamma_E\approx0.5772$ the Euler constant. 
It can be easily checked
that Eq.~(\ref{eq:splitmellin}) reduces to the ordinary splitting functions given 
in~\cite{Dokshitzer:1991wu,Dokshitzer:1978hw} after setting $\ns=1$. 
In the large $z\sim1$ limit, or equivalent large $j\gg1$ we are interested in, the expressions
of the anomalous dimensions (\ref{eq:nuGbis}) and (\ref{eq:nuFbis}) can be re-expressed in the form
\begin{equation}\label{eq:approxdglapaa}
\calp_{qq}(j)\approx4C_FN_s\left(-\ln j+\frac3{4N_s}-\gamma_E\right),\quad
\calp_{gg}(j)\approx4N_cN_s\left(-\ln j+\frac{\beta_0}{N_s}-\gamma_E\right),
\end{equation}
where the asymptotic behavior $\psi(j+1)\approx\ln j$ has been replaced for $j\gg1$.
The Mellin transform can be inverted, with the inversion given by
\begin{equation}\label{eq:Ddelta}
D(x,\Delta\xi)=\frac{1}{2\pi i}\int_Cd j\, x^{-j}{\cal D}(j,\Delta\xi),
\end{equation} 
where the contour $C$ in the complex plane is parallel to the imaginary axis and lies to the right of all 
singularities of ${\cal D}(j,\Delta\xi)$. In (\ref{eq:Ddelta})
$$
\Delta\xi=\frac1{4N_c\beta_0}
\ln\left[\frac{\ln\left(\frac{E\Theta_0}{\Lambda_{QCD}}\right)}{\ln\left(\frac{E\Theta}{\Lambda_{QCD}}\right)}\right].
$$
Integrating (\ref{eq:dglap}) after inserting (\ref{eq:approxdglapaa}) leads to the valence distribution
at large $x\sim1$ \cite{Dokshitzer:1978hw},
\begin{equation}\label{eq:valDAA}
D_{A}^A(x,\Delta\xi)\simeq(1-x)^{-1+4C_AN_s\Delta\xi}\,\frac{\exp[4C_AN_s(\frac3{4N_s}-\gamma_E)\Delta\xi]}
{\Gamma(4C_AN_s\Delta\xi)}.
\end{equation}
where $\beta_0=3/4$ for $n_f=3$ was replaced from (\ref{eq:approxdglapaa})
and the $N_s$ dependence is new. The behavior of the valence distribution (\ref{eq:valDAA}) when $\Theta$
changes can be studied by taking its derivative with respect to
$\xi$. In \cite{Dokshitzer:1991wu}, the valence distribution (\ref{eq:valDAA}) 
was shown to present a maximum at some angle 
$\Theta$ for fixed values of $x$ where the bulk of the jet energy is concentrated. 
Taking $\partial_{\xi}D_A^A=0$ results in,
\begin{equation}\label{eq:digammaeq}
\psi(4C_AN_s\Delta\xi)=\ln(1-x)+\frac3{4N_s}-\gamma_E,
\end{equation}
where here again $\psi(x)$ is the digamma function. 
For $\Delta\xi\to0$, or $\Theta\to\Theta_0$, almost the whole energy will be deposited inside the cone $\Theta_0$
while for $\Theta$ decreasing down to $\Theta\geq E/\Lambda$ the emission outside the cone increases. 
The equation (\ref{eq:digammaeq}) with the $N_s$ dependence is the main result of this paper. It 
should be inverted numerically in order to provide the behavior
of the ratio $\Theta/\Theta_0$ as a function of the energy of the leading parton and the nuclear factor $N_s$ for fixed 
values of $x\sim1$. Symbolically, the inversion can be written in the simple form,
\begin{equation}\label{eq:nscollim}
\frac{\Theta}{\Theta_0}=\left(\frac{E\Theta_0}{\Lambda}\right)^{-\gamma_A(x,N_s)},\quad 
\gamma_A(x,N_s)=1-\exp{\left[-\frac{N_c\beta_0}{C_AN_s}\psi^{-1}\left(\ln(1-x)
+\frac{3}{4N_s}-\gamma_E\right)\right]},
\end{equation}
such that, the collimation is stronger for higher energies in both vacuum and medium jets. As observed in 
(\ref{eq:nscollim}), the $N_s$-dependence contained in the denominator of the exponent should soften the 
collimation of jets produced in heavy ion collisions. Furthermore, 
the function $\gamma_A(x,N_s)$ appearing in the exponent provides the slope of the collimation as a 
function of $x$ and $N_s$. Of course, though (\ref{eq:nscollim}) is model dependent, it may capture the main 
features of a more realistic QCD energy loss picture in heavy ion collisions. For instance, 
it would be interested to extend the vacuum result ((\ref{eq:nscollim}) with $N_s=1$) to the energy loss model 
introduced in \cite{Armesto:2008zza}, which modifies the DGLAP splitting functions by accounting for 
the medium length $L$ and the transport coefficient $\hat q$; however, it stays out of the scope of 
the present paper. 

\section{Phenomenology}
\begin{figure}[h]
\begin{center}
\epsfig{file=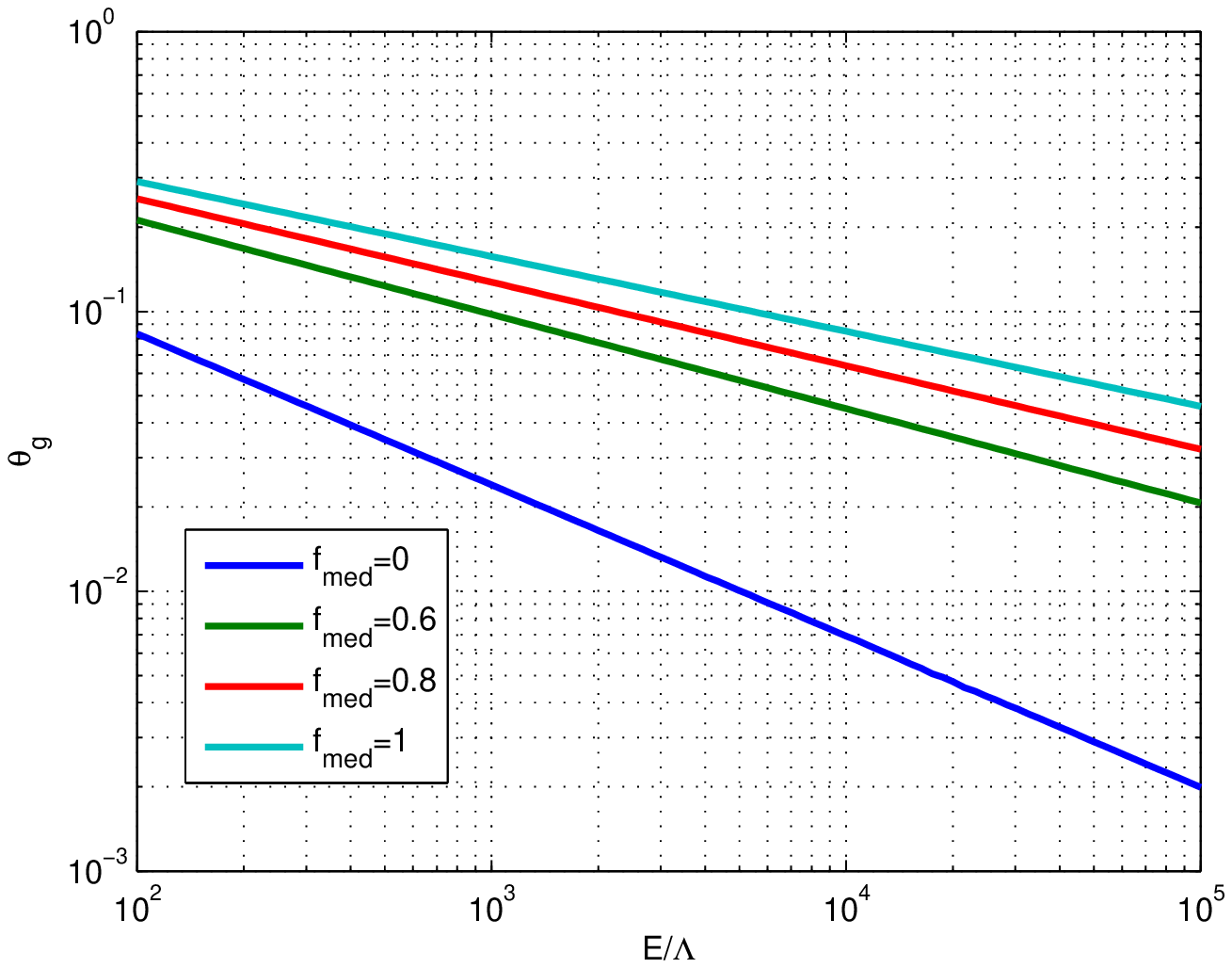, height=6.5truecm,width=7.5truecm}
\epsfig{file=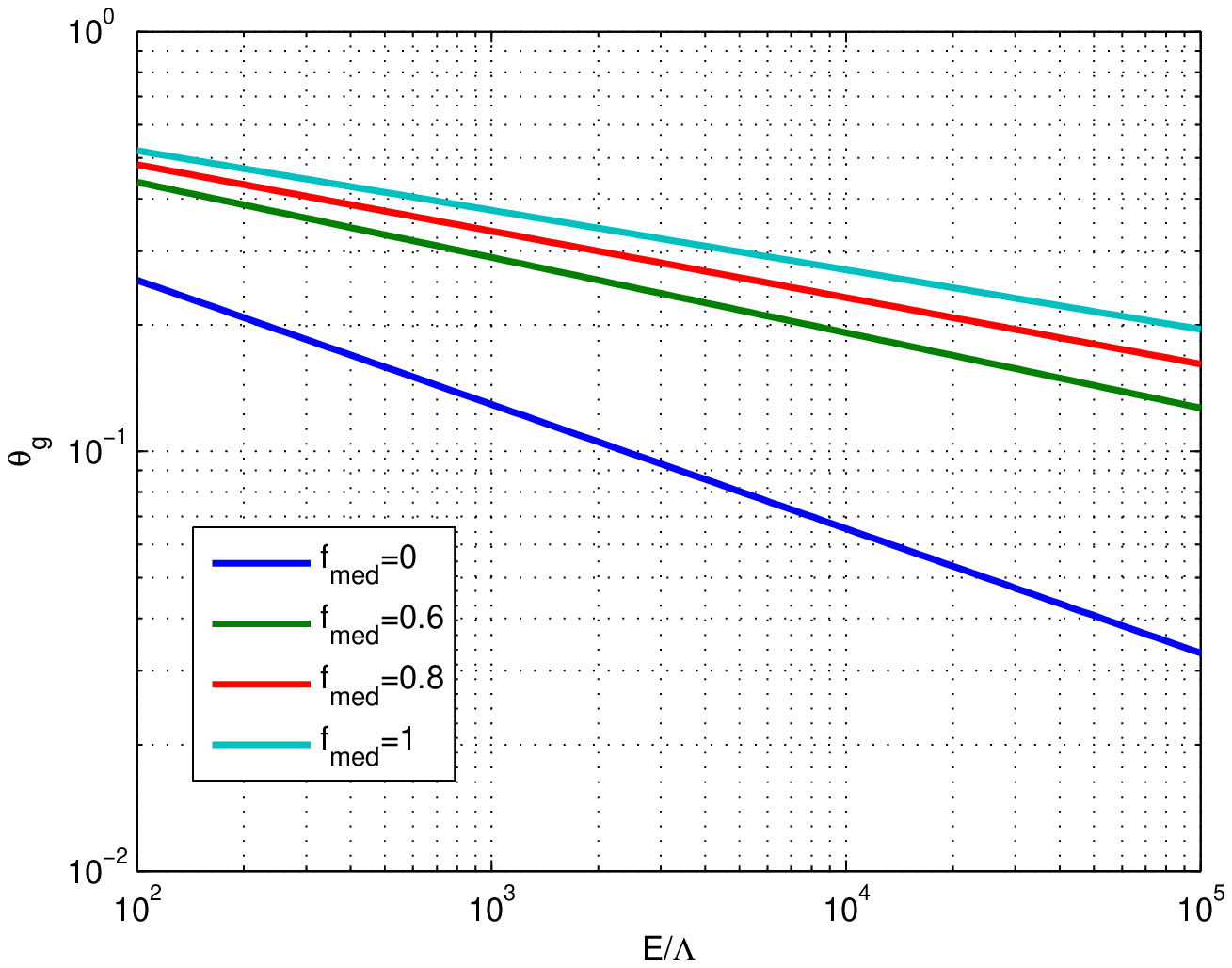, height=6.5truecm,width=7.5truecm}
\caption{\label{fig:gluon} Collimation of energy inside a gluon jet for 
$x=0.5$ (left) and $x=0.9$ (right).}  
\end{center}
\end{figure}

\begin{figure}[h]
\begin{center}
\epsfig{file=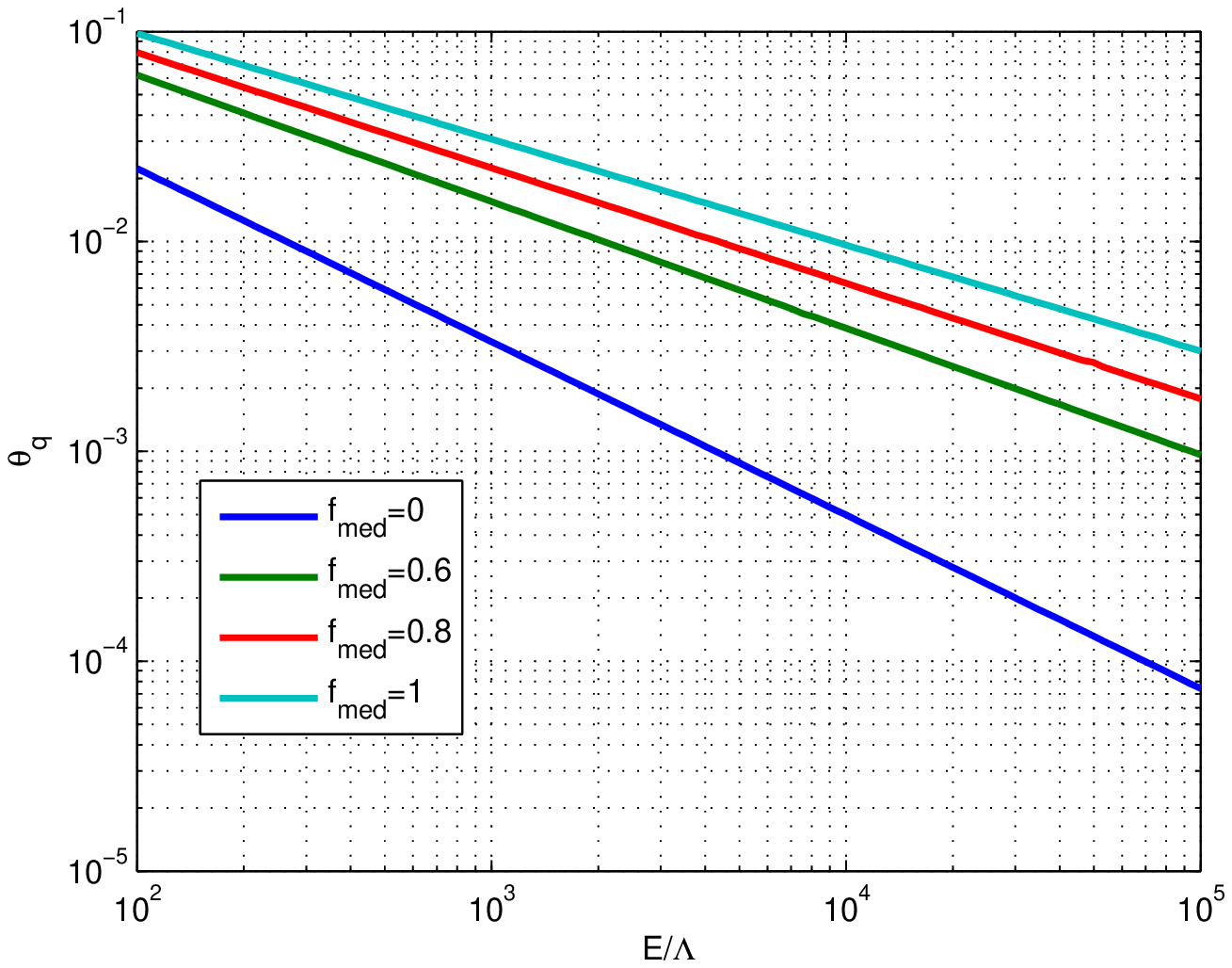, height=6.5truecm,width=7.5truecm}
\epsfig{file=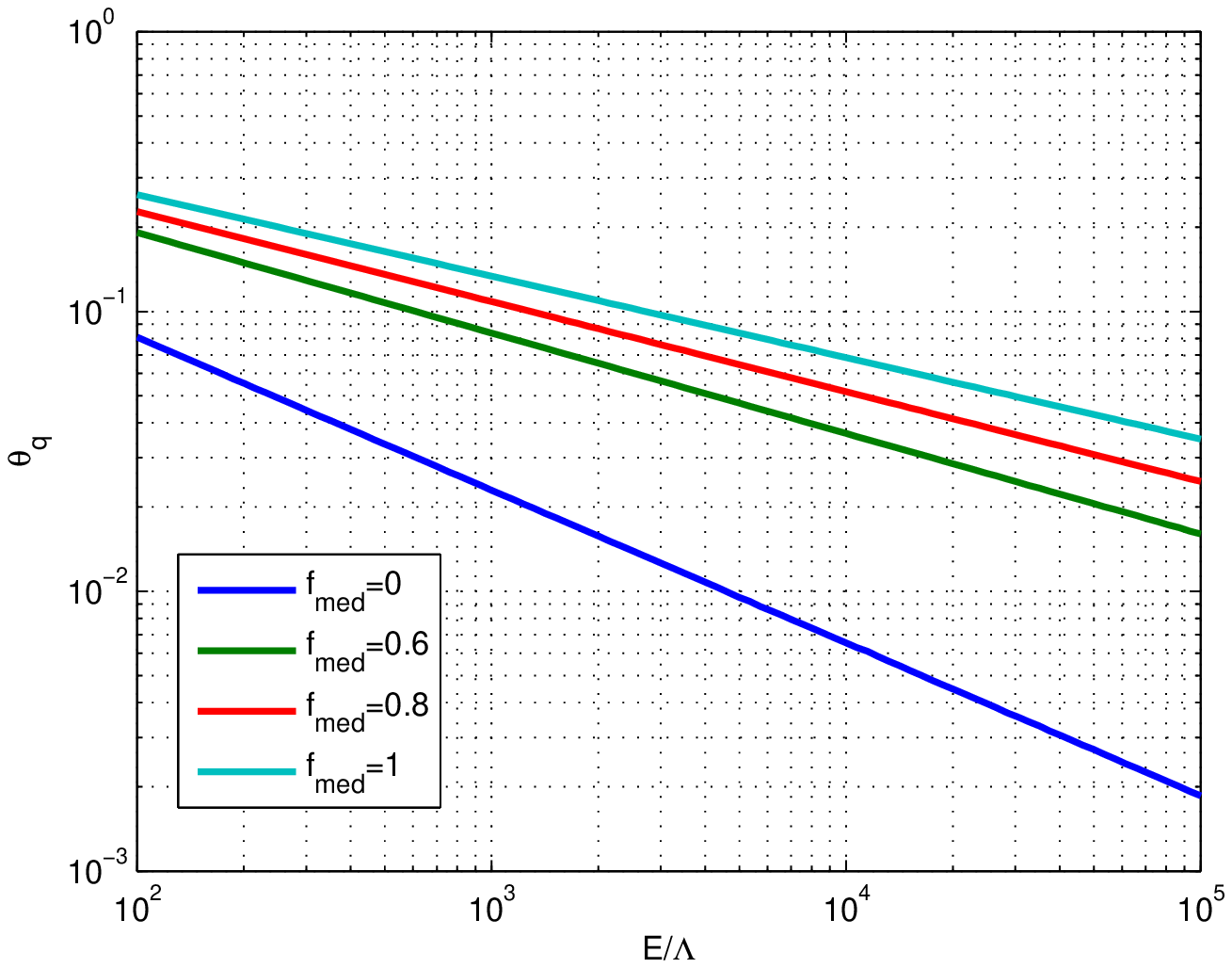, height=6.5truecm,width=7.5truecm}
\caption{\label{fig:quark} Collimation of energy inside a quark jet for 
$x=0.5$ (left) and $x=0.9$ (right).}  
\end{center}
\end{figure}

We display the collimation of energy through the ratio $\frac{\Theta}{\Theta_0}$ from
the numerical inversion of (\ref{eq:digammaeq}). Let us first consider a jet produced in a heavy ion 
collision at energy scales greater than 100 GeV at the LHC and let us set 
$\Theta_0\sim1$ for simplicity. 
For the gluon and quark jets, one sets $C_A=N_c=3$ and $C_A=C_F=4/3$ respectively in (\ref{eq:digammaeq}). 
In figure \ref{fig:gluon} and \ref{fig:quark}, we display the collimation of energy
for two fixed values of energy fractions $x=0.5$ and $x=0.9$ which are in agreement 
with the large $x$ ($x\sim1$) approximation applied in this frame. 
As expected from (\ref{eq:digammaeq}) and observed in figures \ref{fig:gluon} and \ref{fig:quark}, the curves of the 
collimation in the medium ($f_{med}=0.6, 0.8, 1$) are shifted towards higher values of the subjet opening 
angle $\Theta$ as compared with the curve for $\Theta$ in the vacuum. Moreover, the slope of the collimation 
decreases as $f_{med}$ increases. For fixed $\Theta_0$, $E$ and $x$, the same amount of energy contained inside a 
subjet of opening angle $\Theta_1$ in the vacuum is distributed over a broader angular aperture 
$\Theta_2$, such that $\Theta_2>\Theta_1$.
It follows that gluon jets are less collimated than quark jets in the vacuum 
\cite{Dokshitzer:1991wu} and not surprisingly, also in the medium. For $x=0.5$ jets are more collimated
than for higher values of the energy fraction. Finally, it is straightforward
to check that the values of $\gamma_A(x,N_s=1)$ for $x=0.5, 0.9$ are in agreement with the expected values
$\gamma_g(0.5,1)\approx0.54$, $\gamma_g(0.9,1)\approx0.30$, $\gamma_q(0.5,1)\approx0.83$ and
$\gamma_q(0.9,1)\approx0.55$ in the vacuum \cite{Dokshitzer:1991wu,Dokshitzer:1988bq}. 
This analysis could be performed in future measurements by the LHC experiments as a probe for the 
QGP produced in PbPb-collisions.
\section{Summary}

In this paper we have studied the collimation of energy inside jets produced in 
heavy ion collisions. From the production of one jet of opening angle $\Theta_0$ 
initiated by a parton $A$, we considered the production of a subjet $\Theta$ 
($\Theta<\Theta_0$) initiated by a parton $B$, where the definite fraction $x\sim1$ 
of the jet energy is deposited. We made use of the DGLAP evolution equations at large $x$ 
with medium-modified splitting functions, which accounts for medium-induced soft gluon radiation. 
Our study was extended to the distributions $D_A^B(x,Q^2)$ at $x\sim1$ so as to provide
their dependence on the nuclear parameter $N_s$ and therefore, it allowed to determine the behavior 
of the collimation of the jet energy as a function of the same parameter.  

Since these results are model-dependent, more efforts are required towards the construction 
of a more realistic QCD energy loss picture where the medium parameters and medium-induced soft gluon dynamics \cite{MehtarTani:2010ma} could be both taken into account. However, as physically expected from 
the model, jets in the vacuum are more collimated than medium-modified jets and therefore,
there is an evidence for the broadening of jets as gluon radiation is enhanced. 
Quark jets are more collimated than gluon jets in both the vacuum and the medium. 
Therefore, the collimation of energy treated in this paper and the collimation of average multiplicity treated in
\cite{Arleo:2009yu} are both good candidates to investigate the jet quenching.

Finally, in the forthcoming part of this work, we will compare the collimation with
the Monte Carlo in-medium shower YaJEM event generator \cite{Renk:2008pp,Renk:2009nz}
and with preliminary CMS data \cite{Veres}.

\section*{Acknowledgements}

R. P\'erez-Ramos gratefully acknowledges support from ``Centro Nacional de F\'isica de Part\'iculas, 
Astropart\'iculas y Nuclear" (CPAN) under grant FPA2011-23596 and thanks Thorsten Renk for useful 
discussions. Finally, this work is also supported by the Academy researcher program of the Academy 
of Finland, Project No. 130472, where this paper was completed.

%
\end{document}